\documentstyle[prb,aps,multicol]{revtex}
\begin{document}
\draft
\hyphenation{following}
\newcommand{\Eanis}{$E_{anis}$}
\renewcommand{\k}{{\bf k}}
\newcommand{\R}{{\bf R}}
\renewcommand{\r}{{\bf r}}
\newcommand{\spinup}{\left|\uparrow\right>}
\newcommand{\spindown}{\left|\downarrow\right>}
\newcommand{\be}{\begin{equation}}
\newcommand{\ee}{\end{equation}}
\newcommand{\K}{{\bf K}}
\newcommand{\m}{{\bf m}}
\newcommand{\M}{{\bf M}}
%
%
\title{Theory of Magnetocrystalline Anisotropy Energy
for Wires and Corrals of Fe adatoms: A Non-Perturbative Theory}
\author{R. Dru$\check{z}$ini\'c and W. H\"ubner$^{*}$}
\address{Institute for Theoretical Physics, Freie Universit\"at
Berlin, Arnimallee 14, D-14195 Berlin, Germany}
\date{\today}
\maketitle
\begin{abstract}
The magnetocrystalline anisotropy energy 
$E_{anis}$ for free-standing
chains (quantum wires) and rings (quantum corrals) 
of Fe-adatoms $N=$(2...48) is 
determined using an electronic tight-binding theory.
Treating spin-orbit coupling non-perturbatively,
we analyze the relationship between the electronic structure 
of the Fe $d$-electrons
and $E_{anis}(n_{d})$, for both the chain and ring conformations.
We find that $E_{anis}(N)$
is larger for wires than for rings or infinite
monolayers.
Generally $E_{anis}(n_{d})$ decreases in chains upon increasing $N$,
while for rings $E_{anis}(n_{d})$ is essentially independent of $N$.
For increasing $N$, $E_{anis}(n_{d})$ in corrals 
approaches the results for freestanding monolayers.
Small rings exhibit clear odd-even oscillations of $E_{anis}(N)$.
Within our theoretical framework we are able to explain
the experimentally observed
oscillations of $E_{anis}(n_{d})$ during film growth with a period 
of one monolayer.
Finally, a generalization of Hund's
third  rule on spin-orbit coupling to itinerant ferromagnets is proposed. 
\end{abstract}

\begin{multicols}{2}
\section{Introduction}

Over the past decades information technology has been determined
by semiconductor applications.
However, recently the technology of such devices is limited
due to the optical wavelenght. Thus, quantum structures on the 
nanoscale can offer new perspectives
for electronic applications. Manipulations at this scale require
to develop nanotools:
Low-temperature scanning tunneling microscopes allow not only for the
{\em probing} of metallic surfaces with atomic resolution,
but also for single-atom {\em manipulation} on surfaces.
It has recently been demonstrated that lateral structures
of adatoms can be generated by the scanning tunneling
microscope (STM) tip (Crommie {\em et al.}~\cite{Crommie93,Heller94}, 
Meyer {\em et al.}~\cite{Meyer95},
Rieder {\em et al.}~\cite{Rieder96}).
Modern STM-techniques allow now for the preparation of
quantum corrals (QC's). Examples of QC's which have been
prepared are rings of 48 Fe adatoms ~\cite{Crommie93} or 
stadia ~\cite{Heller94}.
The latter shape is motivated by the search for quantum-chaotic
behavior, which so far has not been successful due to strong
scattering of the electrons by the corral atoms.
However the magnetic moment and magnetocrystalline anisotropy
for these nanoscopic structures have not yet been 
determined in experiment.
The QC of adsorbate atoms sets the boundary condition for the
wave function of the substrate surface state. The quantum mechanical
electron distribution of this state can therefore be measured
by STM and is found to be in remarkable agreement with
standard predictions of quantum mechanics.

In this paper, however, we are not interested in the 
surface state but rather in the magnetic anisotropy
properties of the quantum corral itself. Since it is
known that the magnetocrystalline anisotropy energy $E_{anis}$
is enhanced in thin magnetic films compared to 
threedimensional bulk media, one might speculate about a further 
increase of $E_{anis}$ upon reduction of the dimensionality to 1D.
Thus, we calculate $E_{anis}$ for QC's and quantum wires (QW's).
These chains are frequently realized by the enhanced
adsorption probability of adatoms on step edges. 
 
We analyze for chains and rings for different numbers $N$
of Fe-adatoms how the magnetocrystalline anisotropy energy $E_{anis}$ 
depends on the electronic structure. 
To our knowledge, the first attempt of this kind has been made by 
Wang {\em et al.}~\cite{Freeman93} for a pair of two Fe-atoms
to explain the dependence of the electronic states.
While Wang {\em et al.} treated spin-orbit coupling (SOC)
in second-order perturbation theory, we include
SOC completely non-perturbatively
without resorting to degenerate or non-degenerate perturbation theory
of any order. It is shown that for the diatomic pair there are 
important contributions 
to the magnetocrystalline anisotropy energy as a function of the
number $n_{d}$ of $d$-electrons per atom $E_{anis}(n_{d})$ 
due to the different lifting of band degeneracies 
for different magnetization directions .
We find that $E_{anis}\propto\lambda_{so}^2$ is valid
for the diatomic pair.
However, $E_{anis}(n_{d})$ can change its sign with respect to
$\lambda_{so}$ due to level crossings in the energy level scheme 
of the diatomic pair.
In the same way we calculate $E_{anis}(n_{d})$ for chains and rings
with different numbers $N$ of Fe-atoms.
In small rings there are oscillations of $E_{anis}(N)$ between rings
with an even or odd number of atoms.
For our calculation we use two models:\\
(i) an extension of the diatomic pair model~\cite{Freeman93}
to $N$ atoms,\\
(ii) a conventional tight-binding description of the
wires and corrals including SOC.\\
Our paper is organized as follows: In Sec. II, the
diatomic-pair model (II.A) and the tight-binding model (II.B.)  
are presented together with the determination of SOC (II.C)
and $E_{anis}$ (II.D.). 
Results for the diatomic-pair model are shown in Sec. III.A,
while results for the tight-binding model are discussed in
Sec. III.B.
\newpage

\section{Theory}
\subsection{Diatomic-pair model}
We follow the ideas of Wang {\em et al.} and extend their
diatomic pair model to the case of $N$ atoms.  
We start from the following Hamiltonian~\cite{Fletcher52,Slater54}:
\begin{equation}
H=\sum_{i,\gamma,\sigma}\epsilon_{i\gamma} {c_{i\gamma\sigma}^{+}
c_{i\gamma\sigma}} +
\sum_{(i,j)\gamma,\beta,\sigma,\sigma^{'}}t_{i\gamma j\beta}
  {c_{i\gamma\sigma}^{+}c_{j\beta\sigma^{'}}}
\end{equation}
The summation is taken over $N$ sites $i$, while the second sum is over
pairs of nearest neighbors.
As usual $c_{i\gamma\sigma}^{+}$($c_{i\gamma\sigma}$)
creates (destroys) an electron with spin $\sigma$
at site $i$ and orbital $\gamma$.
Furthermore $\epsilon_{i\gamma}$ is the on-site energy
of an electron at site $i$ and orbital $\gamma$. For simplicity
we set $\epsilon_{i\gamma}$=0, since we consider only orbitals
of the 3$d$ type.
The basis functions
$\phi_k$, $k=1,...,5$ ($k=6,...,10$) 
are the atomic 3$d$ orbitals conventionally  denoted by $xy$, $yz$, $zx$, 
$x^2-y^2$ and $3z^2-r^2$, where $z$ denotes the interatomic axis, 
together with the spin eigenstates $\spinup$ 
and $\spindown$ with respect to the spin quantization axis $z_M$.
The complete ferromagnetic tight-binding Hamiltonian is
\begin{equation}
H_{tb}=H+H_{ex},
\end{equation}
with the diagonal on-site matrix $H_{ex}=\sum H_{ex}^{(i)}$ and 
\begin{equation}
H_{ex}^{(i)}=
\left(\begin{array}{cc}
    H_{ex}^{\uparrow\uparrow} &  H_{ex}^{\uparrow\downarrow} \\
    H_{ex}^{\downarrow\uparrow} & H_{ex}^{\downarrow\downarrow} 
     \end{array}\right) 
=\frac{J_{ex}}{2}\left(\begin{array}{cc} -\bf{1}&O\\O&\bf{1}\end{array}\right),
\end{equation}
which is actually independent of $i$. $\bf{1}$ and $\bf{-1}$ 
denote $5\times5$ diagonal matrices, and $O$ is a $5\times5$ zero matrix.
The matrix of the Hamiltonian $H_{tb}$ is then given by:
\begin{equation}
H_{tb}=
\left(\begin{array}{cccccc} H_{ex}^{(1)}&H_{inter}^{(1,2)}&\cdots&\cdots \\
H_{inter}^{(2,1)}&H_{ex}^{(2)}&\cdots&\cdots \\
\vdots&\vdots&\ddots&H_{inter}^{(N-1,N)}\\
\vdots&\vdots&H_{inter}^{(N,N-1)}&H_{ex}^{(N)}
\end{array}\right) 
\end{equation}
with the intersite Hamiltonian matrix
\begin{equation}
H_{inter}^{(i,j)}=:\left(\begin{array}{cc}
H_{inter}^{\uparrow\uparrow}&O\\
O&H_{inter}^{\downarrow\downarrow}
\end{array}\right),
\end{equation}
where
\begin{equation}
H_{inter}^{\downarrow\downarrow}=
H_{inter}^{\uparrow\uparrow}=:\left(\begin{array}{ccccc}
V_{dd\delta}&0&0&0&0\\
0&V_{dd\pi}&0&0&0\\
0&0&V_{dd\pi}&0&0\\
0&0&0&V_{dd\delta}&0\\
0&0&0&0&V_{dd\sigma}
\end{array}\right).
\end{equation}

Due to the symmetry of the atomic 3$d$-orbitals,
the dominant overlaps result from the $d$-orbitals of the same kind,
thus yielding nonvanishing contributions only in the diagonal 
elements of the off-diagonal Hamiltonian  $H_{inter}^{(i,j)}$.
Consequently the nearest neighbor hopping $t_{i\gamma j \beta}$ of Eq. (1)
reduces to $t_{i\gamma j\gamma}$, which is equivalent to the three
independent nonvanishing interactions $V_{dd\sigma}$,
$V_{dd\pi}$, $V_{dd\delta}$ in Eq. (6).
In rings, with periodic boundary conditions, 
the interaction between the first and last atom of the corresponding
chain must be additionally included.
As in ref. [5] we use
$V_{dd\sigma}=-0.25$ eV, 
$V_{dd\pi}=0.18$ eV, $V_{dd\delta}=-0.04$ eV, $J_{ex}=3.0$ meV,
and lattice parameter a=5.98 a.u. which matches the 
fictitious W(001) substrate. 
\subsection{Tight-binding model}
The main disadvantage of the diatomic pair model is its
unrealistic description of tilted $d$-bonds.
To perform more realistic calculations of $E_{anis}(n_{d})$ in rings,
we have to consider how the bond angles are changed between nearest 
neighbors of Fe-atoms.
This aspect is taken into account
in the tight-binding model, within the two-center approximation 
~\cite{Slater54}.
The tight-binding Hamiltonian $H_{tb}$ has the same structure 
for this model as 
the matrix in Eq.(4). Only the elements of $H_{inter}^{(i,j)}$ will
change their values and positions, because the bond
angles have changed. 
The Hamiltonian is parametrized in terms of the two-center
integrals $E_{d,d}(l,m,n)$ of Slater and Koster~\cite{Slater54},
given in Table I, with $d$ as one of the five 3$d$-functions 
$xy$, $yz$, $zx$, $x^{2}-y^{2}$ and $z^{2}$, the latter denoting
$3z^{2}-r^{2}$ for simplicity. 
The direction of the vector $R_{j}-R_{i}$, pointing along 
the bond from one
atom to each of its nearest neighbors, is given by the direction 
cosines $l$, $m$, and $n$. Thus in the tight-binding model both blocks
$H_{inter}^{\uparrow\uparrow}=H_{inter}^{\downarrow\downarrow}$
of $H_{inter}^{(i,j)}$ change to:
\begin{equation}
\left(\begin{array}{ccccc}
E_{xy,xy}&0&0&E_{xy,x^{2}-y^{2}}&E_{xy,z^{2}}\\
0&E_{yz,yz}&0&0&0\\
0&0&E_{zx,zx}&0&0\\
E_{x^{2}-y^{2},xy}&0&0&E_{x^{2}-y^{2},x^{2}-y^{2}}&
E_{x^{2}-y^{2},z^{2}}\\
E_{z^{2},xy}&0&0&E_{z^{2},x^{2}-y^{2}}&E_{z^{2},
z^{2}}
\end{array}\right).
\end{equation}
Since we take the $(x,y)$-plane as the ring or
chain plane, all direction cosines
$n$ must be zero, thus further simplify the nonvanishing two-center
integrals of Table I. The other two direction cosines $l$ and $m$
are determined by the bond angle between the Fe-atoms.

If we calculate the magnetic anisotropy of single or double chains
we assume them to
lie also in the $(x,y)$-plane and being
parallel to the $x$-axis. We couple them either ferromagnetically
or antiferromagnetically, by changing the sign of $J_{ex}$.

\subsection{Spin-orbit coupling}
Magnetocrystalline anisotropy is caused by SOC 
between the $d$-states. SOC is introduced in the usual
spherical on-site
form as $H_{so}=\lambda_{so}{\bf l}\cdot{\bf s}$,
with the orbital and spin moment vectors {\bf l} and {\bf s}
respective and the
atomic spin orbit coupling constant $\lambda_{so}$.
Expressing the components of the orbital momentum operator ${\bf l}$ in 
the rotated frame $(x_M,y_M,z_M)$, where $z_M$
is the spin quantization axis, 
which is parallel to the direction of magnetization
$(\theta,\phi)$, we may rewrite $H_{so}=\sum H_{so}^{(i)}$ in the
following form~\cite{Takayama76,angles}, 
\begin{equation}
H_{so}^{(i)}=:\left(\begin{array}{cc} H_{so}^{\uparrow\uparrow} &
H_{so}^{\uparrow\downarrow} \\ H_{so}^{\downarrow\uparrow} &
H_{so}^{\downarrow\downarrow} \end{array}\right) = 
     \frac{\lambda_{so}}{2}
     \left(\begin{array}{cc}l_{z_M}&l_{x_M}-il_{y_M}\\
           l_{x_M}+il_{y_M}&-l_{z_M},\end{array}\right).
\label{Hso}
\end{equation}
Here is $H_{so}^{\uparrow\uparrow}=-H_{so}^{\downarrow\downarrow}$, and
$H_{so}^{\downarrow\uparrow}=-\left(H_{so}^{\uparrow\downarrow}\right)^*$,
with
\newcommand{\uuu}{u}
\newcommand{\vvv}{v}
\newcommand{\www}{w}
\begin{equation}
\label{Hsoupup}
H_{so}^{\uparrow\uparrow}=\frac{\lambda_{so}}{2}\left(\begin{array}{ccccc}
 0 & i\vvv & -i\uuu & 2i\www & 0 \\
-i\vvv & 0 & i\www &-i\uuu & -\sqrt{3}i\uuu \\
i\uuu & -i\www & 0 & -i\vvv & \sqrt{3}i\vvv \\
-2i\www & i\uuu & i\vvv & 0 & 0 \\
0 & \sqrt{3}i\uuu & -\sqrt{3}i\vvv & 0 & 0
\end{array}\right)
\end{equation}
and
\renewcommand{\uuu}{u'}
\renewcommand{\vvv}{v'}
\renewcommand{\www}{w'}
\begin{equation}
H_{so}^{\uparrow\downarrow}=\frac{\lambda_{so}}{2}\left(\begin{array}{ccccc}
 0 & \vvv & -\uuu & 2\www & 0 \\
-\vvv & 0 & \www &-\uuu & -\sqrt{3}\uuu \\
\uuu & -\www & 0 & -\vvv & \sqrt{3}\vvv \\
-2\www & \uuu & \vvv & 0 & 0 \\
0 & \sqrt{3}\uuu & -\sqrt{3}\vvv & 0 & 0
\end{array}\right).
\end{equation}
The variables $u$, $v$, $w$, $u'$, $v'$, and $w'$ are given by
$u=\sin\theta_M\cos\phi_M$, $v=\sin\theta_M\sin\phi_M$,  
$w=\cos\theta_M$, 
$u'=-\sin\phi_M+i\cos\theta_M\cos\phi_M$, 
$v'=\cos\phi_M+i\cos\theta_M\sin\phi_M$, and
$w'=-i\sin\theta_M$.

The value for the SOC constant $\lambda_{so}$ is taken from the 
Fe-atom~\cite{Argyres55}: $\lambda_{so}=50$~meV.

Now we have set up the Hamiltonian matrix in terms of the magnetization 
direction $(\theta,\phi)$. Since SOC operates to a good approximation
only on-site, the matrix elements of $H_{so}^{i}$ occur exclusively
in the diagonal blocks of $H_{tb}$ in Eq. (4); consequently,
the matrix elements really occur only in the diagonal and off-diagonal
blocks of $H_{ex}^{(i)}$. For chains, $H_{so}^{(i)}$ is independent of
site $i$, while for rings the situation is more complicated.
If the magnetic moments are all parallel, $H_{so}^{(i)}$ is
independent of the site index $i$. 
However, if the magnetic moments have a
radial configuration, $H_{so}^{(i)}$ is dependent on $i$.

Now the total Hamiltonian $H_{tot}$ is
\[ 
H_{tot}=H+H_{ex}+H_{so}.
\]
In our treatment SOC is included non-perturbatively ~\cite{perturb}.
So we get also contributions to $E_{anis}(n_{d})$ from higher
order perturbation theory and may obtain information  about the
scaling behavior of $E_{anis}(n_{d})$ with the SOC constant $\lambda_{so}$.

\subsection{Anisotropy energy}
In both models we define the magnetic anisotropy energy
as the difference of energies for perpendicular
magnetization $E_{\perp}$ and parallel
magnetization $E_{\parallel}$:
\begin{equation}
E_{anis}(n_{d}):=E_{\perp}-
E_{\parallel}.
\label{Eanis1}
\end{equation}
Note that for the diatomic pair-model (as defined in Sec. 2A)
$E_{\perp}$ ($E_{\parallel}$) refers to magnetization
perpendicular (parallel) to the interatomic bonds.
For the tight-binding model (as defined in Sec. 2B) 
$E_{\perp}$ ($E_{\parallel}$), however, refers to magnetization
perpendicular (parallel) to the plane of the QC, which is the 
(x,y)-plane. The QW's in this case are assumed to lie in $x$ direction.
Thus $E_{\perp}$ ($E_{\parallel}$) refers to 
$\M\parallel\hat{z}$ ($\M\parallel\hat{x}$).

We define the magnetic anisotropy energy per atom for 
the diatomic-pair model as:
\begin{equation}
E_{anis}(n_{d}):=E_{tot}(\theta =\pi/2,;n_{d})-
E_{tot}(\theta=0;n_{d}).
\label{Eanis2}
\end{equation}
Due to the chosen coordinate system,
there is no $\phi$-dependence of $E_{anis}(n_{d})$ in contrast
to the tight-binding model.
For the tight-binding model we define $E_{anis}(n_{d})$ as:
\begin{equation}
E_{anis}(n_{d}):=E_{tot}(\theta =0;n_{d})-
E_{tot}(\theta =\pi/2,\phi ;n_{d}),
\label{Eanis3}
\end{equation}
The in-plane angle $\phi$ is choosen 
such that the resulting $|E_{anis}|$ is the largest possible.
$E_{tot}(\theta ,\phi ;n_{d})$ is the ground-state energy 
per atom of the chain or ring with a total of $n_{d}$ $3d$-electrons
per atom, $N$ Fe-atoms and the magnetization direction
denoted by $(\theta,\phi)$  ~\cite{angles}. 
The total energy per atom $E_{tot}$ is given by
\begin{eqnarray*}
 E_{tot}(\theta,\phi;n_{d})&&=\frac{1}{N}
\sum_{m} E_{m}(\theta,\phi)
 f_0\left(E_{m}(\theta,\phi)-E_F(\theta,\phi;n_{d})\right).
\end{eqnarray*}
Here is $f_0(\Delta E)$ the Fermi-function at zero temperature and
$E_F(\theta,\phi;n_{d})$ is the Fermi-energy. For a given
bandfilling $n_{d}$, $E_F(\theta,\phi;n_{d})$ is determined
self-consistently by
\[
n_{d}=\frac{1}{N}
\sum_{m} 
 f_0\left(E_{m}(\theta,\phi)-E_F(\theta,\phi;n_{d})\right).
\]
The $m$-th eigenvalue of the total Hamiltonian $H_{tot}$ 
with magnetization along $(\theta,\phi)$
is given by $E_{m}$.

Note that the magnetic-moment direction enters
only by SOC and can be adjusted for each atomic site individually.
 
Concerning the anticipated magnetic order of the QW and QC 
in our electronic theory we remark the following:
(i) The interatomic distance in our model is chosen in accordance
with the lateral lattice constant of a thin bcc-Fe film on a W(001) 
substrate, which is 
magnetically ordered. However, the STM-experiments produce QC
with the interatomic distance of 8-10 {\AA}, which
probably would not order, except for possible indirect-exchange effects
related to the substrate. Our interest, however, is mainly in small QC's,
which might be produced in the near future, with closer nearest-neighbor
distances and magnetic order.
Furthermore it is important to note that the assumption of 
magnetic long-range
order in our electronic theory is not in contrast 
to the well known general theorems of the (ii) Ising-~\cite{Ising}
and (iii) Heisenberg-models~\cite{Heisenberg}.

\section{Results and Discussion}
\subsection{Results from the diatomic-pair model}
In Fig.~\ref{dimer}, the magnetocrystalline anisotropy energy
$E_{anis}(n_{d})$ per atom is shown as a function of
3$d$-bandfilling $n_d$ for the Fe diatomic pair (dashed curve).
First we note that, as expected, 
$E_{anis}$ is enhanced compared to Fe monolayers 
($E_{anis}=1\ldots16$ meV for the dimer vs. $\approx$ 300 $\mu$eV 
for an Fe monolayer),
because of the symmetry reduction from 2D to 1D.
This is in accordence with recent studies of magnetism in
nickel clusters. Apsel {\em et al.}~\cite{Apsel96} observe
enhanced magnetic moments for small nickel clusters and find
magnetization minima for clusters with closed geometrical
shells and maxima for relatively open clusters.

In Fig.~\ref{dimer}, also the dependence of $E_{anis}(n_{d})$ on the 
scaling of all 
$d$-electron hopping parameters with a overall parameter $t$ for
the diatomic pair is presented (dotted curve).
The absolute value of the magnetic anisotropy energy
$|E_{anis}(n_{d})|$ increases for decreasing 
$t$ (decreasing bandwidth)
for the easy axis perpendicular to the bond ($E_{anis}(n_{d})<0$), 
and decreases
for the easy axis parallel to the bond ($E_{anis}(n_{d})>0$).
Note that, contrary to the situation in thin films, no convergence 
problems occur in $E_{anis}$ due to our finite system. 
Our calculation is performed in real space, and
we sum over discrete energy levels. Thus the change of the
occupation number is $\triangle n_{d}=0.5$, and we get
20 points for $E_{anis}(n_{d})$ of the diatomic pair. 
When the band is completely empty or filled,
no contribution to $E_{anis}$ results: $E_{anis}(n_{d}=0; 10)=0$.
While this is trivially the case for an empty band, this
relation holds with 14 significant digits for the completely
filled case, which yields an independent check of our numerical 
accuracy.  
We find for Fe, with its atomic band filling $n_{d}$=6,
an in plane easy axis for $E_{anis}$(Fe)=6.1 meV. Since
the diatomic pair is a small system, it is not necessary to
assume implicitly hybridization with $s$-electrons.

We find for the diatomic pair a mirror symmetry of $E_{anis}(n_{d})$
with respect to half filling ($n_{d}=5$), where the spin-up subband is 
completely full. However, the corresponding
points of the curve are not totally symmetric, and a close
inspection shows that there are deviations of this mirror symmetry
of the order $10^{-5}$ eV. 
This small symmetry violation results from the combined action
of SOC and hopping interaction.
A detailed analysis shows that this asymmetry increases monotonously 
with increasing SOC.
Especially for $n_{d}$=2.5 and 7.5
this asymmetry is clearly visible in Fig.~\ref{dimer}.
To illustrate the origin of this asymmetry we split the 
spin-orbit coupling matrix $H_{so}$ into two parts
$H_{so}^{par}$ and $H_{so}^{antipar}$, 
for the coupling between
parallel and antiparallel spin states, and recalculate $E_{anis}(n_{d})$ 
with either of the two matrices instead of $H_{so}$ itself:
\[
H_{so}=
\left(\begin{array}{cc} H_{so}^{\uparrow\uparrow} & 0 \\ 0 &
H_{so}^{\downarrow\downarrow} \end{array}\right) + 
\left(\begin{array}{cc} 0 &
H_{so}^{\uparrow\downarrow} \\ H_{so}^{\downarrow\uparrow} &
0 \end{array}\right) =:
H_{so}^{par}+H_{so}^{antipar}.
\] 
So we obtain the curve
$E_{anis}^{antipar}(n_d)$ (solid curve) of Fig.~\ref{dimer},
which exhibits a mirror symmetry 
with respect to half
filling ($n_{d}=5$).
Due to the large exchange splitting $J_{ex}$, which completely
separates the spin subbands, $E_{anis}^{antipar}(n_d)$ is 
very small and thus $E_{anis}\approx E_{anis}^{par}(n_d)$.
The $E_{anis}^{antipar}(n_d)$  contribution 
prefers an easy magnetic
direction perpendicular (parallel) to the molecule axis if the
spin-down subband is less (more) than half-filled, and vice
versa for the majority subband in agreement with the result of 
Wang {\em et al.} [5]. 
Also for larger chains $E_{anis}^{antipar}(n_d)$ 
changes it sign at $n_{d}$=2.5 and 7.5,
which seems to be a general trend also in Fe-monolayers~\cite{Moos}.
This is a generalization of Hund's
third rule on SOC which applies to itinerant 
magnetic systems (see Appendix). 
 
While for Fe parameters the relation 
$E_{anis}^{par}(n_d)+E_{anis}^{antipar}(n_d)
\approx E_{anis}(n_d)$ holds to a good approximation,
small deviations are responsible for the asymmetry mentioned above.
This asymmetry only occurs if we take into account
at least the matrix elements of $H_{so}$
for the $\pi$- and $\delta$-bonds and is absent for all other
combinations.

The electronic origin of $E_{anis}(n_{d})$ of the diatomic pair
can be explained from its energy level scheme in Fig.~\ref{lev}, for
the two magnetization directions $E_{m}$($\theta$=0) 
(parallel to molecule axis) 
and $E_{m}$($\theta$=$pi$/2) (perpendicular to molecule axis).
Only the spin-up states are shown in Fig.~\ref{lev} for simplicity;
due to the exchange splitting $J_{ex}$=3.0 eV the spin-down states 
(not shown in the figure) are located symmetrically to
these bands with respect to the zero of the energy scale. 
The individual levels of the spin-up subband
have been labeled by the numbers 1$\ldots$10 in the absence of SOC.
Without SOC, the
lowest bonding (level 1) and the highest antibonding states (level 10)
result from the splittings caused by $V_{dd\sigma}$. 
Due to the inequalities $|V_{dd\sigma}|>|V_{dd\pi}|>|V_{dd\delta}|$ 
(Sec. II.B) the lowest bonding- and antibonding states are split by
$2|V_{dd\sigma}|$, while in each case two doubly degenerate
$\pi$-($\delta$-)bonding and antibonding states result from
$V_{dd\pi}$ and $V_{dd\delta}$ (see Eq. 6).
Thus the levels 2 and 3 (4 and 5) of Fig.~\ref{lev} correspond to the 
$\pi$-($\delta$-)bonding states, while the levels 8 and 9 (6 and 7) are
equivalent to the corresponding antibonding states.
These remaining degeneracies
are lifted very strongly by SOC for $E_{m}$($\theta$=0) and
therefore level crossings may occur.
For $E_{m}$($\theta$=$pi$/2) the degeneracies are only weakly lifted.
The electronic origin of $E_{anis}(n_{d})$ for
the diatomic pair results from
(i) different SOC-induced $shifting$ $of$ $occupied$, $nondegenerate$ $levels$
for the two magnetization directions, and
(ii) $different$ $lifting$ $of$ $degenerate$ $levels$.
Note that the lifting of degeneracies can favor a magnetization
parallel as well as perpendicular to the bond axis.

Next we investigate the dependence of $E_{anis}$ on $\lambda_{so}$.
We find $E_{anis}(n_d)\propto\lambda_{so}^2$ 
for all of the $n_d$ values of the diatomic pair 
in agreement with Wang {\em et al.}\/~\cite{Freeman93}, irrespective
of the value of the spin-orbit coupling constant. To demonstrate this, 
we show in Fig.~\ref{soc} the magnetocrystalline anisotropy energy 
$E_{anis}(\lambda_{so})$ for the bandfillings
$n_{d}$=1 and $n_{d}$=2.5.
In Fig.~\ref{soc}, however, we see a discontinuous change of the slope
for $n_{d}$=2.5, because a level crossing (see Fig.~\ref{lev})
has occurred between 
the levels 5 and 6 for $\lambda_{so}$=0.04 eV. 
The inversion of these levels is indicated
by the labeling in Fig.~\ref{lev}, for $\lambda$=0.05 eV.
Thus $E_{anis}(n_{d})$ can change its sign
because of the level crossings.
To explain for $n_{d}$=1 the dependence of $E_{anis}$ on SOC,
we have to study the contribution of level shifts for different
magnetization directions.
We find that the shifting of $nondegenerate$ $states$ is for both
magnetization directions  proportional to $\lambda_{so}^2$. For the
case of an 
perpendicular easy axis, the prefactor is larger.
The shifting of $degenerate$ $states$ for magnetization
parallel to the bond axis is in all cases proportional to  $\lambda_{so}$ and
for magnetization perpendicular to the bond axis proportional
to $\lambda_{so}^2$.
To illustrate this we show in the inset of Fig.~\ref{soc} the dependence of 
the level shifts on 
$\lambda_{so}$ for the nondegenerate level 1 (this corresponds to
$n_{d}$=0.5) and the degenerate level 2 ($n_{d}$=1), for
both magnetization directions ($\theta=0; \pi/2$) for the diatomic pair.
For $\theta=\pi/2$ and small SOC there is a flat slope of the 
energy of both levels, while the slope increases with increasing 
$\lambda_{so}$. Otherwise, the slope of the energy of both levels for
$\theta=0$ is increasing less strongly with increasing $\lambda_{so}$.   
Thus in Fig.~\ref{soc}  the magnetic anisotropy energy $E_{anis}(n_{d}=1)$ 
decreases for larger SOC with increasing $\lambda_{so}$, as a result
of the different position of the energy levels for
the two magnetization directions entering the calculation of 
$E_{anis}(n_{d})$. 
This yields another reason for the
sign change of $E_{anis}(n_{d})$.

An other important consequence of the
levelcrossing can be seen in Fig.~\ref{dimer}. The 
magnetic anisotropy energy $E_{anis}(n_{d})$
of the quarter-band filling ($n_{d}$=2.5 and 7.5)
of both subbands depends sensitively on
the hopping parameters, because the level crossings then 
occur for smaller values of $\lambda_{so}$
(decreasing $t$) or, in the other case of increasing $t$ for
larger values of $\lambda_{so}$.

In Fig.~\ref{fring} (dashed curve), $E_{anis}(n_{d})$ is shown 
as a function of 3$d$-band filling $n_d$ for the five-atom Fe ring. 
In the five-atom ring we find a translational 
symmetry of $E_{anis}(n_{d})$
with respect to half filling. This symmetry behavior of
$E_{anis}(n_{d})$ is found in all rings 
with a small odd number $N$=(3,7,9,11,..) of adatoms. 
In rings, with an even number of atoms and all chains irrespective on
whether $N$ is odd or even, we find a mirror symmetry of $E_{anis}(n_{d})$
with respect to half filling. This behavior is also valid in
the tight-binding model and will be discussed again in Sec. III.B.
If the number of atoms continues increasing ($N$=45,47,..), 
we then find a mirror symmetry of $E_{anis}(n_{d})$
with respect to half filling also in odd rings.

\subsection{Results of the tight-binding model}
The tight-binding model is better than the diatomic-pair model,
because it additionally allows to take into account the following points :
(i) The change of the binding angles between nearest neighbors
of adatoms in rings is considered in a realistic way.
(ii) Due to the extra $\phi$-dependence of
$E_{anis}(n_{d})$ in the tight-binding model, it is possible to 
choose any magnetic-moment direction in the plane of the ring or chain.
(iii) Furthermore, in the tight-binding model it is possible to
couple the free-standing chains and rings to the substrate.
This extension of the theory, however, has not been
performed in this work.

For all chains it is easily checked that 
$E_{anis}(n_{d})$ is equal in both models.
There is
a general trend that $E_{anis}(N)$ decreases with increasing
number $N$ of Fe atoms in chains (compare Figs. 1 and 8). 
For rings this is not the case
which will be discussed later.

The solid curve in Fig.~\ref{fring} shows $E_{anis}(n_{d})$
for the tight-binding five ring.
Although this curve is different from the result in the 
diatomic pair-model, the translational symmetry survives.
A detailed analysis yields that the tight-binding model has the same
symmetry behavior for small odd rings as the diatomic-pair model. 
The translational symmetry can be traced back to the absence of a
mirror symmetry between the levels in each of the two spin 
subbands with respect to $\pm J_{ex}/2$: 
In odd rings and chains both spin-subbands contain 
an odd number of electron states. If there the degeneracies of both subbands
resulting from the exchange splitting are all lifted due to the
hybridization, as is the case for odd rings, it is not possible 
to find a mirror symmetry in each of
the two subbands. Since both spin-subbands have a translational
symmetry with respect to the zero of the energy scale,
$E_{anis}(n_{d})$ has in this case also a translational 
symmetry with respect to half
bandfilling. In all odd chains always five degeneracies (angular
momentum algebra of 3$d$-states)
are not lifted by hybridization. The remaining even number of
the lifted degeneracies has a mirror symmetry
around $\pm J_{ex}/2$ in both subbands, thus
$E_{anis}(n_{d})$ has also a mirror symmetry with respect to half
bandfilling.
In larger odd rings the translational symmetry of $E_{anis}(n_{d})$
is lost. Due to the increased number of levels in both subbands
the additional lifting of the five degeneracies, by closing of the
corresponding odd chain to the ring, will not immediately destroy 
the mirror symmetry in both spin-subbands, as is the case for
small odd QC's.
  
A very important point for the theory of small particles is the
existence of shell structure.
Here we discuss in particular the oscillations of $E_{anis}$ as
a function of bandfilling $n_{d}$
for rings
with an even or odd number of adatoms.
If the magnetic moment is
along the layer normal, it lies in planes perpendicular
to all bonds of the ring. But if the magnetic moment
lies in the ring plane (assuming exchange-enforced parallel
moment alignment), it is $twice$ parallel to
a bond axis in rings with an even number of $N$ and $once$ in rings with 
an odd number of $N$. This is schematically 
illustrated in Fig.~\ref{figoe2} for the trimer (5(a)), tetramer (5(b)),
and hexamer ring 5(c)). 
In even rings, where $N$ is a multiple of four the moment lies also
twice perpendicular to a bond axis (see Fig. 5(b)).

We investigate exemplarily for the four-Fe atom ring
three different configurations of the magnetic moment direction,
which are shown in Fig.~\ref{vring}. Since
in larger rings the
exchange interaction could possibly not enforce 
parallel alignment of all magnetic moments as 
assumed previously~\cite{ferrofl},
we calculated $E_{anis}(n_{d})$ for 
parallel magnetic moment alignment (p),
radial moment configurations along the bonds (i) or
half of the nearest-neighbor bond angle (h).
Since there is a $\phi$-dependence of $E_{anis}(n_{d})$ in the 
tight-binding model, 
we find the following in-plane symmetry in rings for the first two 
configurations (p) and (i): 
\begin{equation}
E_{anis}(\theta;\phi=0;n_{d})=E_{anis}(\theta;\phi=360/N;n_{d}).
\end{equation}
Note this symmetry is absent for the case (h).
In Fig.~\ref{vring}, $E_{anis}(n_{d})$ is shown 
as a function of 3$d$-bandfilling $n_d$ of the four atom 
Fe ring, for the three configurations (p), (i), and (h).
Especially for the $\pi$-bond, the magnetic anisotropy energy
$E_{anis}(n_{d})$ has the lowest values for the configuration
(p).
However, this is not the ground state configuration for Fe
with atomic bandfilling $n_{d}$=6.
For all three configurations we get a parallel 
easy axis of $E_{anis}$(Fe), while its value is the largest 
where the magnetic moments are arranged in
the configuration (h).
This is the ground state of the three configurations for Fe, 
with the lowest total energy.
In order to determine the ground-state
configuration for the case of bandfilling around $n_{d}$=7.5, 
we have to notice that
we get a perpendicular easy axis for the two radial configurations
and an in-plane easy axis for the configuration (p). 
In this case both
radial configurations have the same total energy, which has to
be compared with the total energy that results from the
magnetization in-plane (p). A close inspection for all discrete
bandfillings between $n_{d}$=7 and $n_{d}$=8
shows that the configuration (p) has the lowest total energy 
in this range.

Within our theoretical framework we are able to explain a recently
discovered phenomenon,
viz. oscillations of $E_{anis}(n_{d})$ during film growth with
a period of one monolayer ~\cite{Fassbender95}.
The in-plane lattice spacing during epitaxial growth of Co 
on a Cu(001) single-crystal substrate is found to
oscillate as a function of coverage
and produces anisotropy-oscillations~\cite{privat96}. 
These oscillations are schematically illustrated in Fig.~\ref{Anisosz},
which shows the connection between the change of the lattice spacing
from $R$ to $R_{1}$ due to the growth and the
corresponding change of $E_{anis}$.
After starting the deposition, Co islands nucleate on
the Cu surface. The Co atoms are relaxing in the direction
of the center of the island. This gives rise to a reduced 
in-plane lattice spacing.
If the growth of the N+1-layer is completed, $E_{anis}$(N+1-layer)
nearly returns to the value of $E_{anis}$(N-layer).
For uncompleted layers, however, the in-plane lattice constant is very
different from that of complete layers and therefore  
oscillations occur in $E_{anis}$(N-layer) as expected.
As a model assumption of this situation the four atom Fe-ring
can be considered. 
We scale all $d$-electron hopping parameters belonging to $one$ bond of
the four-atom Fe ring with a parameter $t'$, while leaving
the others unchanged. 
Then we find also anisotropy-oscillations, due to the changed position
of some levels in the energy level scheme of the four atom Fe ring.
For decreasing $t'$ for one bond, $E_{anis}(n_{d})$
is increasing or decreasing dependent on $n_{d}$.

Finally we would like to discuss in this paper some results
for larger QC's and QW's, in order to reach  the experimental situation. 
In Fig.~\ref{tbr48}, $E_{anis}(n_{d})$ is shown 
as a function of 3$d$-bandfilling $n_d$ of the 48 atom 
Fe ring (solid curve), for the p-configuration of Fig.~\ref{vring}. 
A comparison of the three configurations
(p, i, h) shows only marginal deviations of the
total energy for the 48 QC, thus we may assume the p-configuration 
to be the ground state. 
For an effective bandfilling of $n_{d}^{eff}$=6.6 we find an 
in-plane easy axis with
$E_{anis}$(Fe)=1.45 meV, assuming
implicitly a hybridization of the $d$-electrons with 
$s$-electrons for this larger system. 
The shape of the curve is very similar to that
of the tight-binding calculations for free-standing 
Fe-monolayers~\cite{Moos,Moos95}.
As can be seen from the comparison of Figs.~\ref{tbr48}, ~\ref{fring},
and ~\ref{vring},  
$E_{anis}(N)$ is  essentially independent of $N$.

Since metal atoms
aggregating during film growth preferably on step edges
(see ref.[20] for Fe/W(110)) may cause one-dimensional
metallization and new magnetic properties,
it is of interest
to calculate the magnetocrystalline anisotropy of larger QW's.
The origin of magnetic step anisotropies results from the modified local 
symmetry of step atoms compared to surface atoms.
In Fig.~\ref{tbr48}, $E_{anis}(n_{d})$ is shown 
as a function of 3$d$-bandfilling $n_d$ for the 48-atom 
Fe chain, for two different scalings of the hopping parameters
(the dashed curve refers to the unscaled case $t$=1).
The value of $E_{anis}(n_{d})$ is considerably lower compared to the
diatomic pair. However, it is still much larger then the values for
freestanding Fe-monolayers. 
The shape of the curve for $t$=1 is very similar to
the diatomic pair (see Fig.~\ref{dimer}),
irrespective
of atomic number $N$. Thus we expect 
very large contributions to $E_{anis}(n_{d})$
in all chains from SOC-induced lifting of degeneracies.
Comparison of Figs.~\ref{dimer} and ~\ref{tbr48}
indicates that, upon decreasing $t$, the shape of the curve 
will change similarly for the 48 atom QW and for the dimer
due to levelcrossings.
 
In order to show how sensitive the the magnetocrystalline anisotropy 
energy depends on spin configuration,
we calculate $E_{anis}(n_{d})$  
for two 48-atom 
Fe chains which are antiferromagnetically (AF) or ferromagnetically (FM)
coupled to each other (Fig.~\ref{chain48a}). 
In these 96 atom systems $E_{anis}(N)$
has further decreased in both cases.  
In AF coupled chains $E_{anis}(n_{d})$
is less dependent on the scaling of the interchain hopping than in  
FM coupled chains.
Note that the shape of $E_{anis}(n_{d})$ for AF coupled chains is similar to 
that of the 48 atom Fe chain. In this case the shape
will not change if the interchain hopping $tl$ is varied,
in contrast to FM coupled chains.
Thus the breakdown of the
one dimensionality in two FM coupled chains will change
$E_{anis}(n_{d})$ very strongly. We find in this case a perpendicular 
easy axis with $E_{anis}$(Fe)=$-0.36$ meV (tl=1).
This value is of the same order of magnitude as for
Fe-monolayers. Experimental results show~\cite{Albrecht92} that 
surface and step anisotropies are of the same order of
magnitude, both exhibiting symmetry-breaking 
as was introduced by N\'eel's phenomenological model~\cite{Neel54,Neel53}.
This is in agreement with our results for
two FM coupled chains. The opposite sign and same order of magnitude
for surface and step anisotropies is reproduced.

\section{Summary and Outlook}
A calculation of the magnetocrystalline anisotropy
energy $E_{anis}(n_{d})$ for free-standing
chains and rings of Fe adatoms is performed. 
SOC is included non-perturbatively.
$E_{anis}(N)$ for 1D chains is larger then for 2D rings, but
in both cases $E_{anis}$ is enhanced copared to Fe-monolayers.
Generally, very large contributions to $E_{anis}(n_{d})$
result from the SOC-induced lifting of degeneracies.
The shape of $E_{anis}(N)$ is in all chains very similar.
In rings $E_{anis}(N)$ is essentially independent of $N$, while in chains
$E_{anis}(N)$ generally decreases upon increasing $N$. For all chains
and even rings, there is a mirror symmetry of $E_{anis}(n_{d})$,
while for small odd rings we find a translational symmetry.
In small rings, there are odd-even oscillations of $E_{anis}(N)$,
which might be of technological relevance 
for magneto-optical switching.
Larger rings approach the results for freestanding Fe-monolayers.
$E_{anis}(n_{d})$ is found to be more sensitive to interchain 
hopping variation for two ferromagnetically than for two 
antiferromagnetically coupled chains. 
A generalization of Hund's third rule on SOC is found.
Oscillations of the magnetocrystalline anisotropy energy due the
growth of Co on Cu have been explained.

Our results indicate that low-temperature scanning tunneling 
microscopy is a promising new tool for the preparation of
atomic-scale nanostructures. These nanostructures exhibit
interesting magnetocrystalline anisotropy properties which
can be tailored by choosing appropriate atomic species and atom
number, ring or chain conformations, and a suitable substrate.
It will be of theoretical interest to incorporate the substrate
degrees of freedom
in the investigation of the magnetocrystalline anisotropy of 
quantum wires and corrals.

It is then to be expected to establish many of the phenomena
successfully demonstrated in thin film magnetism, also for
one-dimensional magnetic systems, such as enhanced moments,
giant magnetoresistance, quantum well oscillations,
giant nonlinear Kerr rotation, or magnetic reorientation.

\section*{ACKNOWLEDGMENT}
We would like to thank Prof. K. H. Bennemann
for many stimulating discussions and his continued support of this
work.

\begin{appendix} 
\section*{Hund's third rule.}
The curve of $E_{anis}^{antipar}(n_{d})$ changes it sign at
$n_{d}$=2.5 and 7.5 (see Fig.1).
This sign change occurs exactly, when in each case the two subbands
are half filled.

Wang {\em et al.}~\cite{Freeman93} also found this behavior for
SOC between the opposite spin states for the dimer which
can be understood from the following equation ($z$
denoting the symmetry axis and $x$ an arbitrary direction
in the perpendicular plane),
\begin{eqnarray}
\Delta E_{ud}&=&E_{ud}(x)-E_{ud}(z)\nonumber\\ 
&=&\frac{\xi^2}{\Delta E_{ex}}\sum_{o^-}<o^-|\frac{1}{2}
   (3L_{z}^{2}-L^{2})|o^->.
\end{eqnarray}
\end{appendix}
$ud$ ($o^{-}$) represents SOC between the opposite spin states
(represents occupied spin-down states), and $\Delta E_{ex}$
is the exchange splitting.
The sign of $\Delta E_{ud}$ depends only on the axial component of the
angular momentum. Three states ($L_{z}=0,\pm 1$), which get
first occupied, give
negativ contributions and two ($L_{z}=\pm 2$) contribute positively.
Due to the hybridization, the
two antibonding states with $L_{z}=\pm 2$, which contribute positively,
will be occupied next, and therfore it comes to a sign change.

This behavior is also valid in larger chains and 
monolayers~\cite{Moos,Moos95}.
If one spin-subband is half filled, there is $L$=0 and
from this results no contribution to the anisotropy energy.
The antibonding states will be occupied in reverse order
and thus there is a sign change at $n_{d}$=2.5 and 7.5.

\newpage 
\begin{figure}
\caption{Dependence of the magnetocrystalline anisotropy
  energy $E_{anis}$ on the 3$d$-bandfilling $n_{d}$ for
  a diatomic pair. While the dashed curve refers to the hoppings given
  in the text the dotted curve refers to hoppings scaled with an
  overall factor of $t$=0.7.
  Negative values of $E_{anis}(n_{d})$ correspond to a perpendicular
  easy axis.
  The solid curve show the contributions of
  $E_{anis}^{antipar}(n_{d})$  to $E_{anis}(n_{d})$ from
  antiparallel spins. As can be seen the main contribution 
  to $E_{anis}(n_{d})$ results from 
  $E_{anis}^{par}(n_{d})$, due
  to the large exchange splitting $J_{ex}$ in Fe.
  The zeros in $E_{anis}^{antipar}(n_{d})$ at $n_{d}$=2.5
  and 7.5 indicate a sign change inposed by a generalization of Hund's
  third rule (see text).}
\label{dimer}
\vspace{.5cm}
\caption{Energy levels of the diatomic pair for the two 
  magnetization directions $E_{m}(\theta=0)$ (magnetization in 
  bond direction) and $E_m(\theta=\pi/2)$
  (perpendicular to the bond axis).
  For simplicity only the spin-up states of the
  energy scale are shown in the plot.
  For magnetization perpendicular to the bond axis the degeneracies are very
  weakly lifted by SOC, in contrast to the case of magnetization in 
  the bond direction, where SOC lifts the degeneracies so strongly, that
  level 5 and 6 intersect.}
\label{lev}
\vspace{.5cm}
\caption{The dependence of the magnetocrystalline anisotropy
  energy $E_{anis}$ on $\lambda_{SO}$ (eV) for $n_{d}$=2.5
  (solid line) and 1 (dashed line) for
  the diatomic pair. For $n_{d}$=2.5 a discontinuous change of the
  slope occurs due to the crossing of levels 5 and 6 in the case
  of a magnetization parallel to the bond axis.
  For $n_{d}$=1 $E_{anis}$ decreases with increasing $\lambda_{so}$.
  This behavior is
  explained in the inset of Fig. 3 (see text).
  The inset shows the shift of levels 1 ($n_{d}$=0.5) and 2 
  ($n_{d}$=1) as a function
  of $\lambda_{SO}$  for the diatomic pair and
  magnetization in bond direction ($\theta=0$) and perpendicular to bond
  ($\theta=\pi/2$).}
\label{soc}
\vspace{.5cm}
\caption{The dependence of the magnetocrystalline anisotropy
 energy $E_{anis}$ on the 3$d$-bandfilling $n_{d}$ for
 the five atom ring calculated in the diatomic pair model (dashed curve)
 and the tight-binding model (solid curve).
 In both models, there is a translational symmetry with respect 
 to half filling $n_{d}=5$. 
 In the tight-binding model all moments in-plane are choosen to be 
 parallel to each other for this case.}
\label{fring}
\vspace{.5cm}
\caption{Odd-even oscillations of the rings 
  (schematic).}
\label{figoe2}
\vspace{.5cm}
\caption{Dependence of the magnetocrystalline anisotropy
  energy $E_{anis}$ on the 3$d$-bandfilling $n_{d}$ for
  a four atom ring calculated in the tight-binding model
  for the three displayed magnetic moment configurations.
  In (p) all moments are parallel to each other, while
  in (i) and (h) the two radial configurations are given, where all
  moments are in the bond directions or half to these.}
\label{vring}
\vspace{.5cm}
\caption{Schematic sketch of anisotropy oscillations during the 
  growth of Co on Cu. We expect that the change of the lattice
  spacing from $R$ to $R_{1}$ due the growth of the (N+1)-layer 
  will produce
  oscillations of $E_{anis}$, which are found in the calculation
  for the four atom Fe ring, used to simulate this system.}
\label{Anisosz}
\vspace{.5cm}
\caption{Dependence of the magnetocrystalline anisotropy
  energy $E_{anis}$ on the 3$d$-bandfilling $n_{d}$ for
  a 48 atom ring (solid curve) for the tight-binding model, with the
  configuration that all moments in plane are parallel to each other.
  The dashed (dotted) curves shows the dependence of $E_{anis}(n_{d})$ 
  for a 48 atom chain for two different scalings of the hopping
  t=1 (t=1/3).}
\label{tbr48}
\vspace{.5cm}
\caption{Dependence of the magnetocrystalline anisotropy
  energy $E_{anis}$ on the 3$d$-bandfilling $n_{d}$ for
  two antiferromagnetically (dashed curve)
  and ferromagnetically (solid and dotted curve) coupled 48-atom 
  Fe chains for the tight-binding model.}
\label{chain48a}
\end{figure}
\begin{table}
\caption{The nonvanishing energy integrals in terms of two-center
integrals, which we used for the tight-binding model.}
\centering
\begin{tabular}{l@{}cr}
$E_{xy,xy}$ & $3l^{2}m^{2}(V_{dd\sigma})+(l^{2}+m^{2}-4l^{2}m^{2})
(V_{dd\pi})+(n^{2}+l^{2}m^{2})(V_{dd\delta})$ \\
$E_{yz,yz}$ & $3m^{2}n^{2}(V_{dd\sigma})+(n^{2}+m^{2}-4m^{2}n^{2})
(V_{dd\pi})+(l^{2}+m^{2}n^{2})(V_{dd\delta})$ \\
$E_{zx,zx}$ & $3n^{2}l^{2}(V_{dd\sigma})+(n^{2}+l^{2}-4l^{2}n^{2})
(V_{dd\pi})+(m^{2}+l^{2}n^{2})(V_{dd\delta})$ \\
$E_{xy,x^{2}-y^{2}}$ & ${\frac{3}{2}}lm(l^{2}-m^{2})(V_{dd\sigma})+
2lm(m^{2}-l^{2})(V_{dd\pi})+{\frac{1}{2}}lm(l^{2}-m^{2})(V_{dd\delta})$ \\
$E_{xy,z^{2}}$ & $\sqrt{3}lm[n^{2}-\frac{1}{2}(l^{2}+m^{2})]
(V_{dd\sigma})-2\sqrt{3}lmn^{2}(V_{dd\pi})+
\frac{1}{2}\sqrt{3}ml(1+n^{2})(V_{dd\delta})$ \\
$E_{x^{2}-y^{2},x^{2}-y^{2}}$ & ${\frac{3}{4}}(l^{2}-m^{2})^{2}
(V_{dd\sigma})+[l^{2}+m^{2}-(l^{2}-m^{2})^{2}](V_{dd\pi})+[n^{2}+
{\frac{1}{4}}(l^{2}-m^{2})^{2}](V_{dd\delta})$ \\
$E_{x^{2}-y^{2},z^{2}}$ & $\frac{1}{2}\sqrt{3}(l^{2}-m^{2})
[n^{2}-{\frac{1}{2}}(l^{2}+m^{2})](V_{dd\sigma})+\frac{1}{4}\sqrt{3}
(1+n^{2})(l^{2}-m^{2})(V_{dd\delta})$ \\
$E_{z^{2},z^{2}}$ & $[n^{2}-\frac{1}{2}(l^{2}+m^{2})]
^{2}(V_{dd\sigma})+3n^{2}(l^{2}+m^{2})(V_{dd\pi})+\frac{3}{4}
(l^{2}+m^{2})^{2}(V_{dd\delta})$ \\
\end{tabular}
\end{table}
\newpage
\newcommand{\PRL}{Phys. Rev. Lett.~}
\newcommand{\PRB}{Phys. Rev. B~}
\newcommand{\PR}{Phys. Rev.~}

\end{multicols}

\begin{thebibliography}{99}
\bibitem{angles} The angle $\theta$ denotes the angle between the 
magnetization direction $z_M$ and the molecule axis or surface 
normal $\hat{z}$. 
$\phi$ is the angle between the $x$-axis and the projection of $z_M$
in the plane of perpendicular magnetization or in the
plane of the monolayer.
\bibitem{Ising} It is well known that the 1D Ising-model,
which is applicable for QW's and QC's (with periodic boundary conditions)
exhibits a phase transition only at $T$=0.
However, in our case we may implicitly assume hybridization with
the substrate, which introduces some 2D-character to our system.
\bibitem{Heisenberg} The 2D Heisenberg model should exhibit
no phase transition if no anisotropy is  involved; the
latter, however, is the subject of this work. Thus the assumption of
magnetic long range order seems justified.
\end{thebibliography}
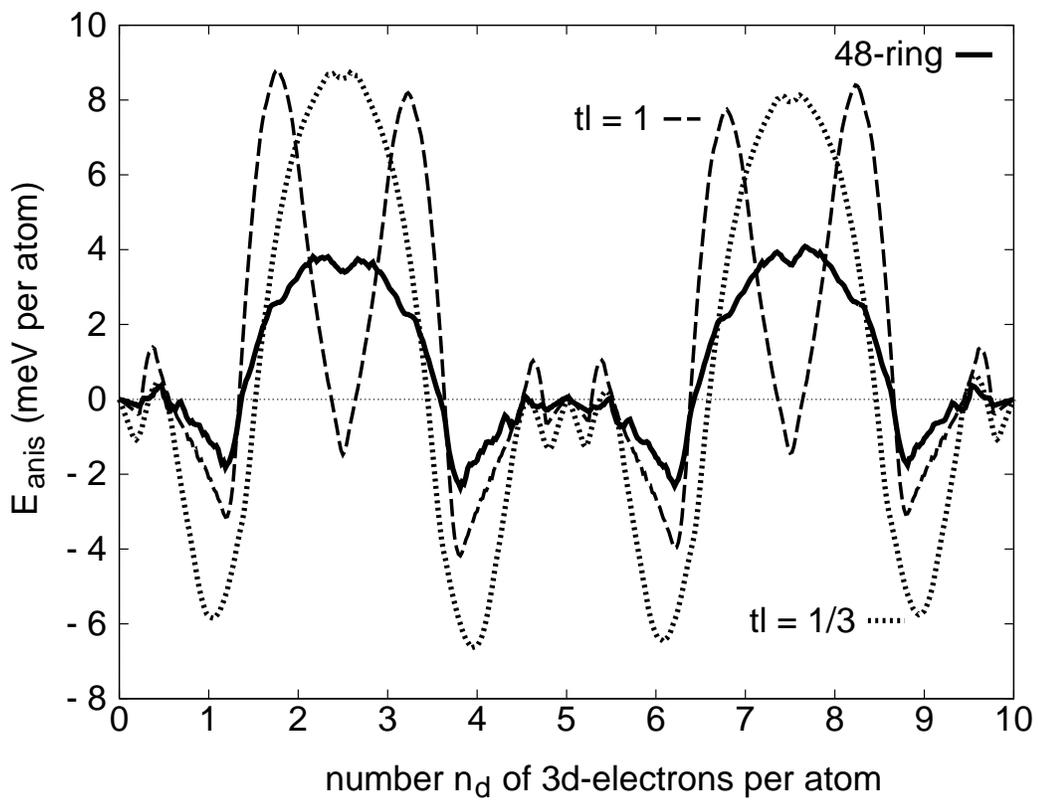
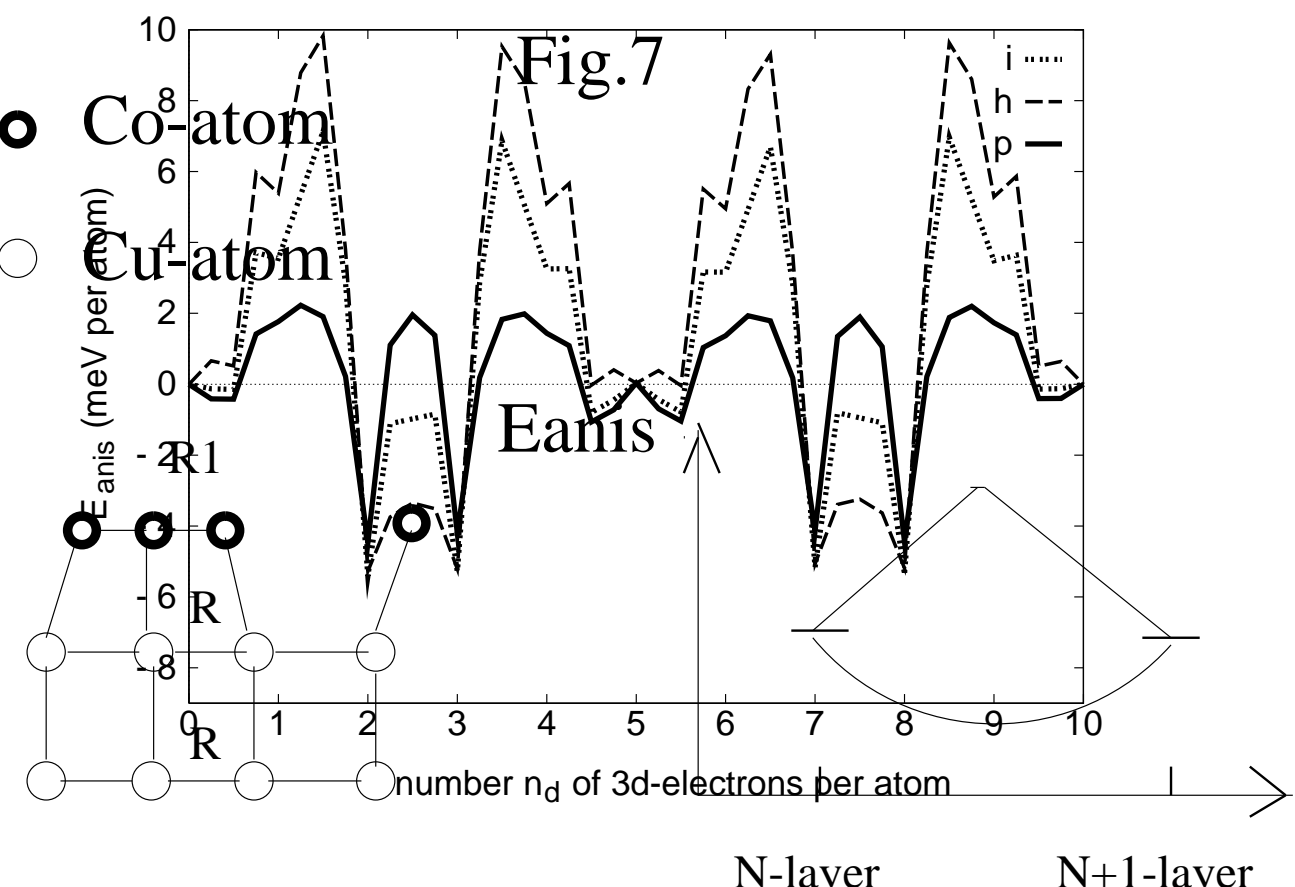
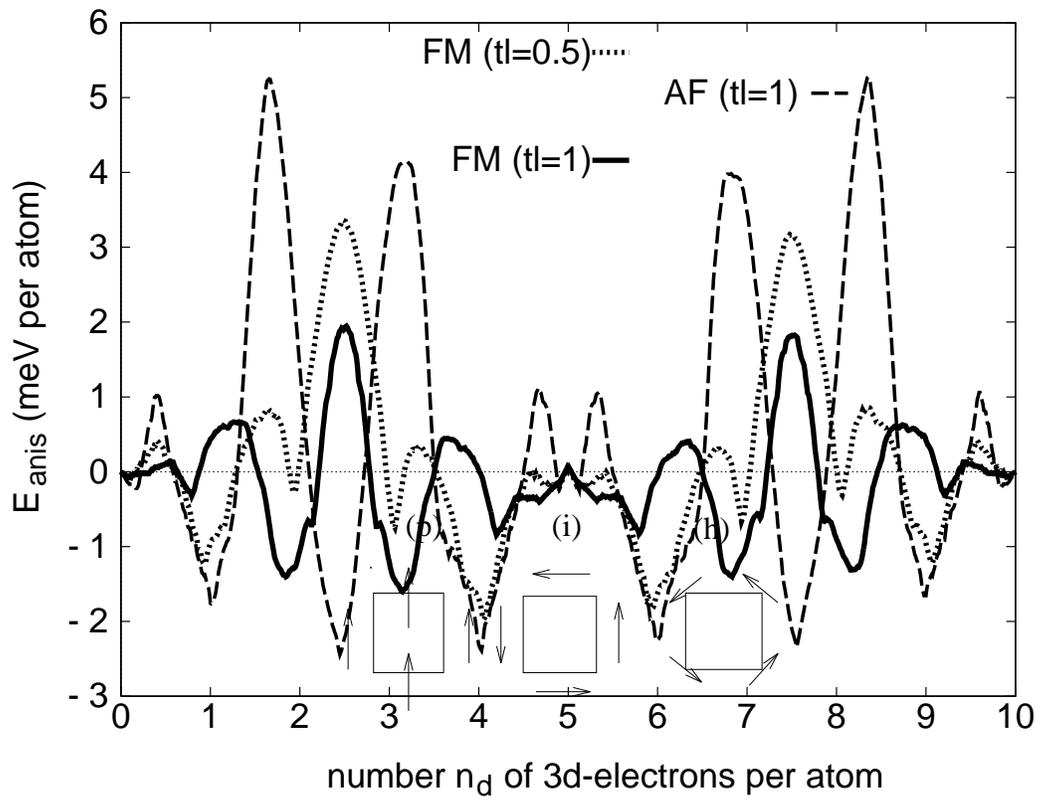
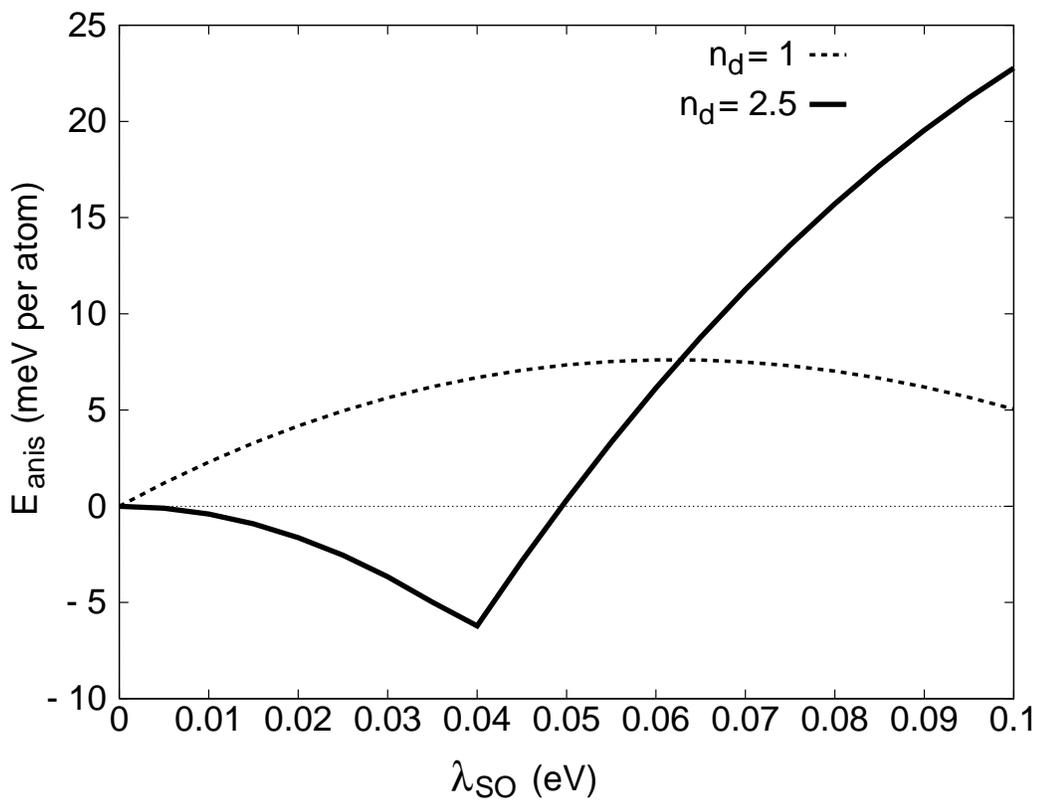
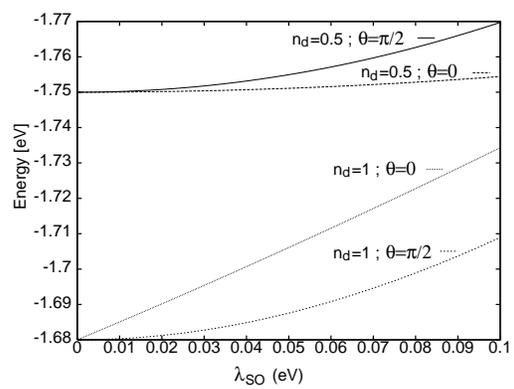
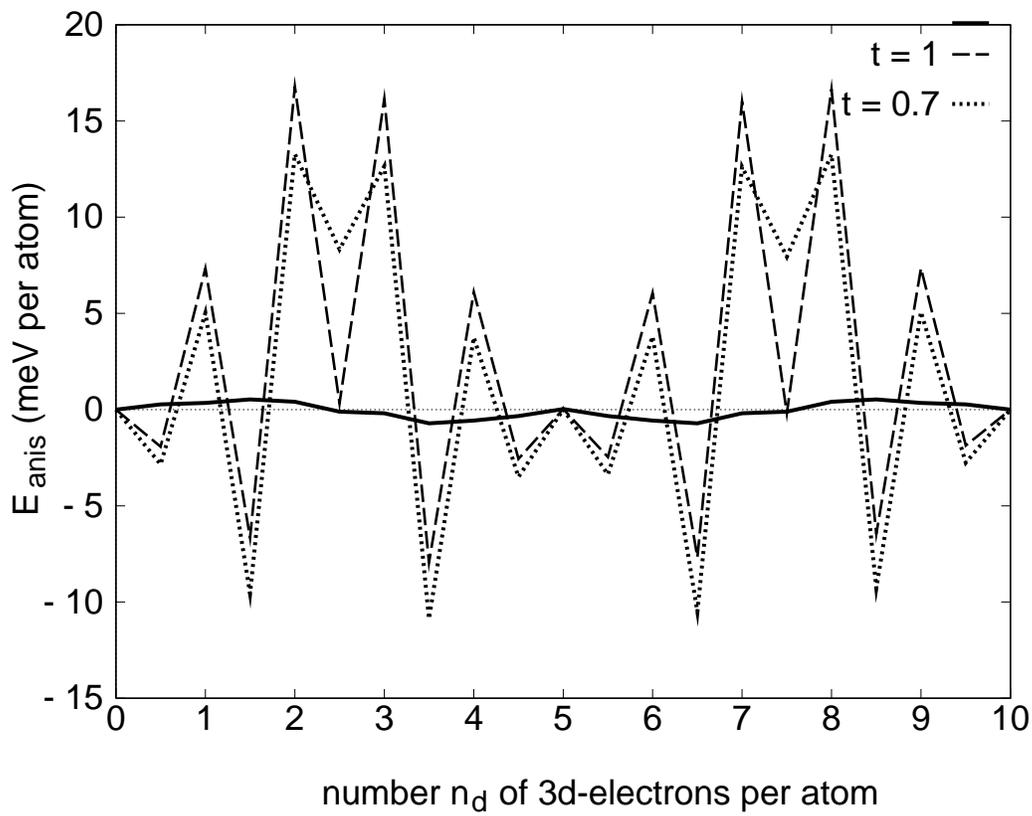
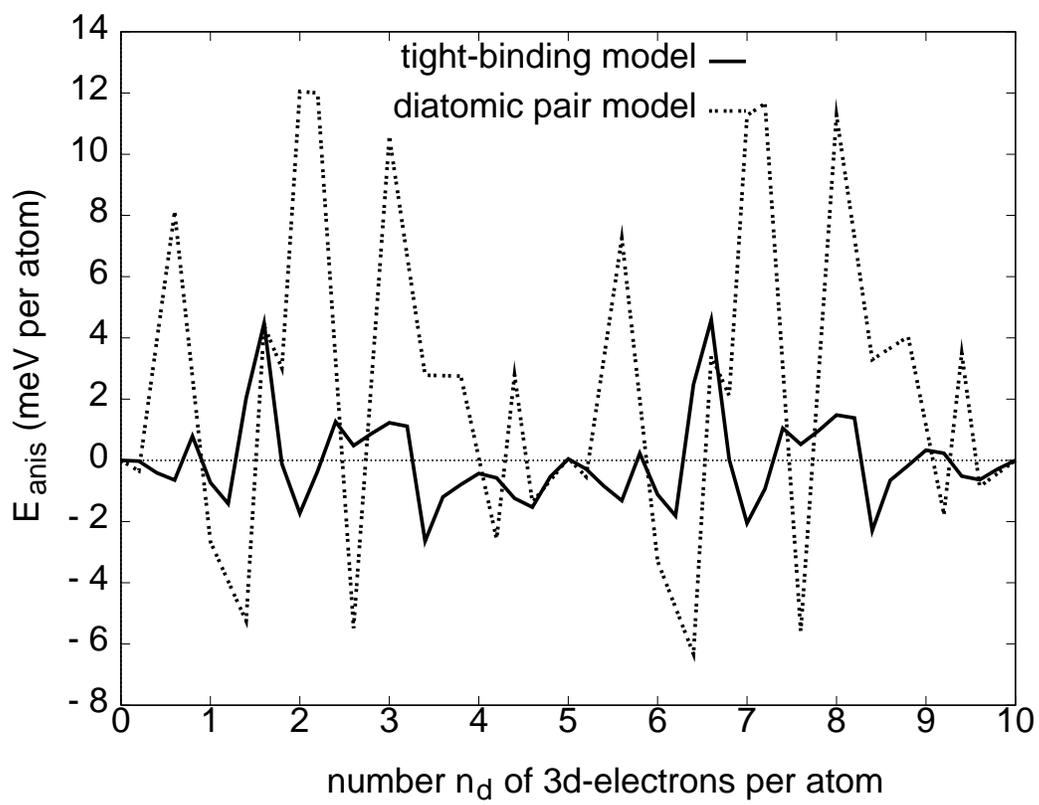
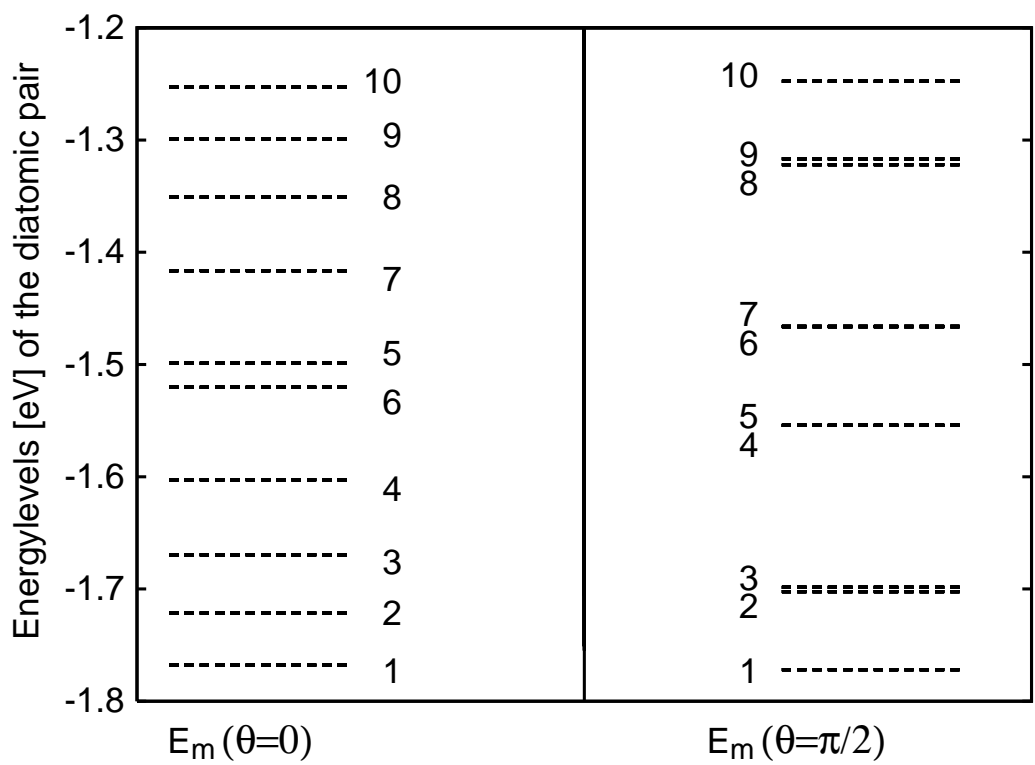
\end{document}